\documentclass{article}

\usepackage[preprint]{styling}
\usepackage[pdftex]{graphicx}
\usepackage{makecell}

\usepackage[utf8]{inputenc} 
\usepackage[T1]{fontenc}    
\usepackage{hyperref}       
\usepackage{url}            
\usepackage{booktabs}       
\usepackage{amsfonts}       
\usepackage{nicefrac}       
\usepackage{microtype}      
\usepackage{xcolor}         

\title{Hitchhiking Rides Dataset: Two decades of crowd-sourced records on stochastic traveling}

\author{%
  Till Wenke\thanks{Hitchhiking since 2021. Personal website: https://tillwenke.github.io/} \\
  Independent Researcher\\
  Hitchwiki.org\\
  Berlin, Germany\\
  \texttt{lastname.firstname[at]gmail.com} \\
}

\begin{document}

\maketitle

\begin{abstract}
Hitchhiking, a spontaneous and decentralized mode of travel, has long eluded systematic study due to its informal nature. This paper presents and analyzes the largest known structured dataset of hitchhiking rides, comprising over 63,000 entries collected over nearly two decades through platforms associated with \texttt{hitchwiki.org} and lately on \texttt{hitchmap.com}. By leveraging crowd-sourced contributions, the dataset captures key spatiotemporal and strategic aspects of hitchhiking. This work documents the dataset’s origins, evolution, and community-driven maintenance, highlighting its Europe-centric distribution, seasonal patterns, and reliance on a small number of highly active contributors. Through exploratory analyses, I examine waiting times, user behavior, and comment metadata, shedding light on the lived realities of hitchhikers. While the dataset has inherent biases and limitations—such as demographic skew and unverifiable entries it offers a rare and valuable window into an alternative form of mobility. I conclude by outlining future directions for enriching the dataset and advancing research on hitchhiking as both a transportation practice and cultural phenomenon.\footnote{Dataset published at: \url{https://huggingface.co/datasets/Hitchwiki/hitchhiking-rides-dataset}}

\end{abstract}

\section{Introduction}\label{sec:intro}

Hitchhiking is a form of transportation in which individuals solicit free rides from passing vehicles, typically by standing at the roadside and signaling to drivers. As defined in the \citet{garrett_hitchhiking_2014}, “Hitchhiking as a form of transportation involves the solicitation of free rides from passing motor vehicles at the side of a road, normally from strangers (although hitchhiking can occasionally involve other forms of vehicles)”.

In the context of this work and the dataset that is introduced, I extend the conventional definition to include any form of spontaneous ride solicitation — whether by private car, van, boat, or other mode of transport. While the majority of rides recorded still involve private automobiles, this broader framing acknowledges the diversity of transport modes encountered by hitchhikers around the world.

A defining characteristic of hitchhiking is its inherent spontaneity and uncertainty. Unlike scheduled forms of transport, hitchhiking is stochastic by nature: the time until pickup, the destination of the ride, and the route taken are all unpredictable. This uncertainty, while often perceived as a risk, also imbues hitchhiking with a unique flexibility and appeal. It is precisely this stochastic aspect that motivates the systematic collection and analysis of crowd-sourced hitchhiking ride data.

The creation of the dataset presented in this work has been made possible by several technological and infrastructural developments over the past two decades. The rise of the internet and collaborative platforms such as \texttt{hitchwiki.org}\footnote{\url{https://hitchwiki.org/en/Main_Page}}, which is based on MediaWiki, provided a foundation for knowledge exchange and community building among hitchhikers. In parallel, the proliferation of smartphones enabled real-time ride tracking and GPS-based location data collection, significantly increasing the precision and utility of the dataset over time.

\begin{figure}[!h]
    \centering
     \includegraphics[width=0.8\linewidth]{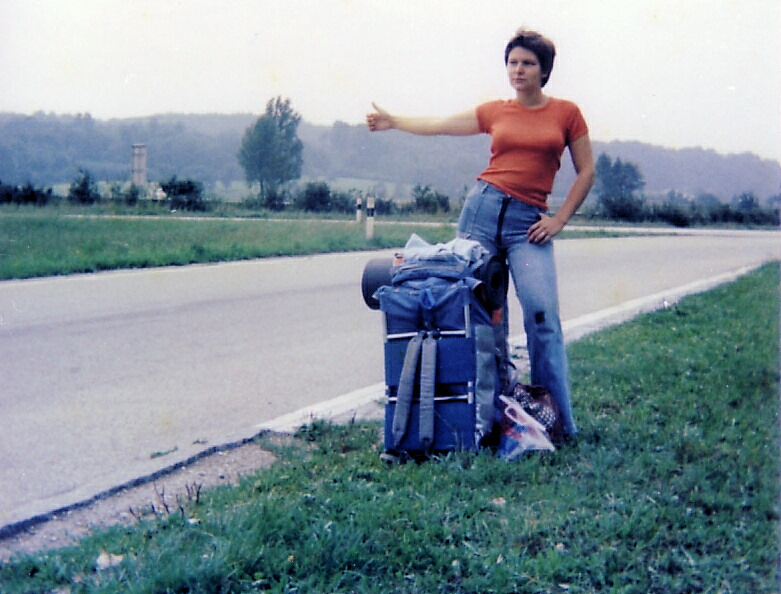}\\[1ex]
    \caption{Symbolic picture of a hitchhiker. Image taken in Luxembourg, 1977. Source: \url{https://hitchwiki.org/en/File:Hitchhiker-Luxemburg-1977.jpg}}
    \label{fig:hitchhiker}
\end{figure}

This paper introduces and analyzes what is, to the best of my knowledge, the first and largest structured dataset of world-wide hitchhiking rides. The aim is not only to provide insights into hitchhiking as a practice, but also to demonstrate the potential of collaborative, decentralized data collection in documenting and sustaining alternative modes of transportation.

\section{Related Work}\label{sec:related}

The dataset presented in this work is, to the best of my knowledge, the first large-scale dataset on hitchhiking that combines spatial and temporal information collected over a span of nearly two decades. In contrast, most prior studies on hitchhiking have focused on isolated aspects of the phenomenon, often within narrowly defined geographic areas, fixed time frames, or specific social experiments.

Early efforts to quantify hitchhiking success can be found in experimental studies such as \citet{crassweller_experimental_1972}, which attempted to test assumptions about the factors influencing whether a hitchhiker gets picked up. While these experiments were pioneering, they were limited in scope and scale.

More recently, a meta-analysis by \citet{kotz_base-rate_2017} synthesized findings from various small-scale studies and individual reports. This work revealed clear gender-based differences in hitchhiking success rates and was based largely on experimental or observational setups where only potential rides were recorded, rather than actual rides taken. For example, \citet{gueguen_color_2012} used staged scenarios to observe driver behavior, without following through on real-world travel outcomes. In contrast, the dataset presented in this paper consists exclusively of actual hitchhiking rides, providing a more accurate representation of real-world hitchhiking experiences.

The long-term and crowd-sourced nature of this dataset also enables finer-grained spatiotemporal analysis, opening the door to deeper insights into regional, seasonal, and infrastructural influences on hitchhiking success, something previously only addressed in isolated studies. For instance, \citet{vedernikov_hitchhikers_2016} examined spatial factors such as road type and the presence of facilities, using the state of this dataset that was available at the time.

An effort to support hitchhiking through route planning algorithms has been proposed  by\citet{vedernikov_hitchhikers_2017}, though it neglects regional variations in attitudes and hitchhiking viability. The dataset presented in this paper could significantly enhance such models by incorporating real-world, user-generated data over a wide spatial and temporal range.

Additionally, the visualization of hitchhiking data as in \cite{sulyok_average_2019} and \cite{wenke_where_2024} has evolved alongside this dataset, with new methods being developed to display trends across geography. These visual tools have both informed the community and contributed to the public perception and documentation of hitchhiking as a legitimate mode of travel.

Overall, while previous work has laid important theoretical and empirical foundations, this dataset introduces an unprecedented opportunity for data-driven research into hitchhiking behavior at a global and long-term scale.

\section{Collection Methodology}\label{sec:methodology}

As of the time of writing, the dataset contains 63,165 recorded rides, of which 7,240 do not include a timestamp. In this section I want to provide an overview about the history of it and how the changing sites that collected the data shaped it.

The dataset was primarily collected and shared through changing single central platforms associated with \texttt{hitchwiki.org}, a collaborative project focused on the culture and practice of hitchhiking. With this, the purpose of tracking hitchhiking rides has always also been its visualization on a map to give hitchhikers a point of information beyond the textual content that is provided through the MediaWiki.

\begin{figure}[!h]
    \centering
     \includegraphics[width = 1.0\textwidth]{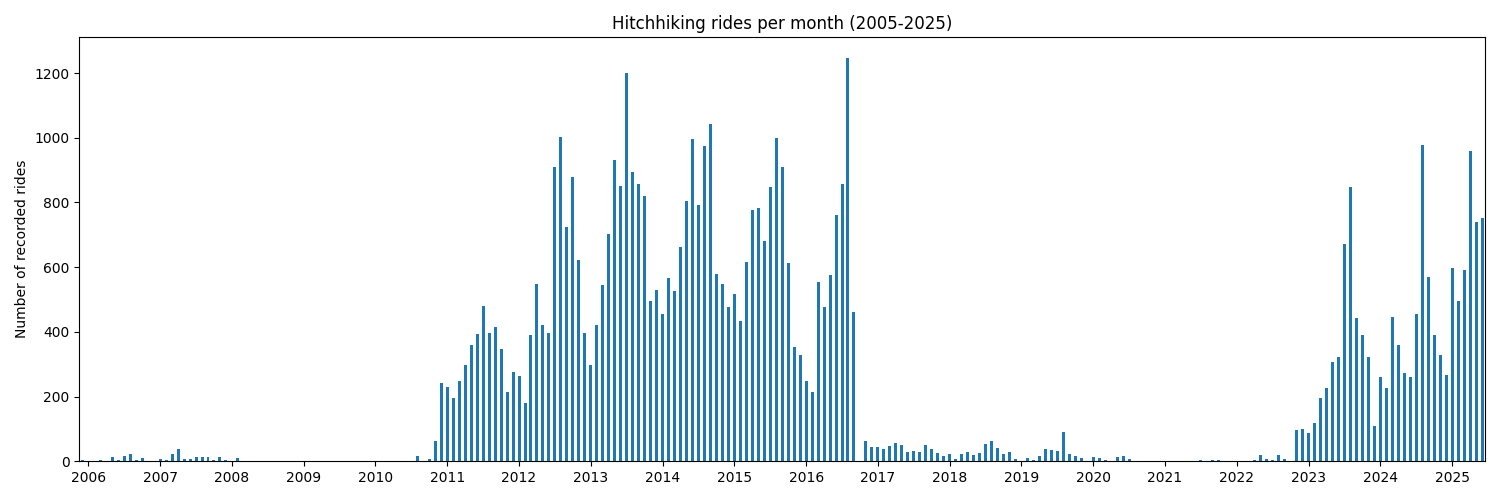}\\[1 ex]
    \caption{Distribution of recorded rides over the course of collection of this dataset.}
    \label{fig:rides_per_month_2005_2025}
\end{figure}

Initially, hitchhiking-related information on \texttt{hitchwiki.org} was limited to city-specific spot descriptions, starting in late 2005. Although informative, these entries did not fully capture the diversity or temporal dimension of individual hitchhiking trips. The first attempt to jointly log hitchhiking rides digitally began in early 2005 on \texttt{hitchbase.com}\footnote{\url{https://web.archive.org/web/20050715000000*/hitchbase.com}} (it also served as a motivation to start \texttt{hitchwiki.org} as a similar but MediaWiki-based solution) which was run by the German hitchhiking association Abgefahren e.V. There hitchhikers collected textual experience reports with rough descriptions of their starting point and the direction of their ride. The lack of precise starting locations is the reason that this historic artifact of the online hitchhiking community does not find its way into this dataset. Quite soon, in December 2005, the website \texttt{liftershalte.info}, a small project that was used by a handful of dedicated hitchhikers, began to log hitchhiking rides in a more structured form, which laid the foundation for this dataset. This marks an important shift in comparison to the studies in \ref{sec:related} that are mostly manually conducted in a fixed set of spots.

\begin{figure}[!h]
    \centering
     \includegraphics[width = 0.2\textwidth]{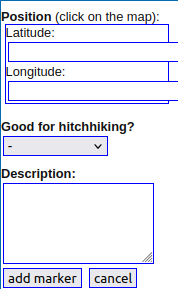}\\[1 ex]
    \caption{Form to input hitchhiking-ride-related information on \texttt{liftershalte.info}}
    \label{fig:liftershalte_form}
\end{figure}

\begin{table}
  \label{tab:liftershalte_form}
  \centering
  \begin{tabular}{ll}
    \toprule
    \cmidrule(r){1-2}
    \makecell{Feature prompt}     & \makecell{Input}     \\
    \midrule
    \makecell{Position (click on the map) Latitude Longitude}    & \makecell{[float, float]} \\
    \midrule
    \makecell{Good for hitchhiking?} & \makecell{1 (excellent), 2 (good), 3 (average), 4 (bad), 5 (senseless)} \\
    \midrule
    \makecell{Description}     & \makecell{string} \\
    \bottomrule
  \end{tabular}
  \caption{Prompts to ask for feature  and their possible input values on \texttt{liftershalte.info}}
\end{table}

On \texttt{liftershalte.info} features of a hitchhiking ride were collected through the form shown in \ref{fig:liftershalte_form} and fields presented in \ref{tab:liftershalte_form} could be collected. Although it cannot be tracked down how \texttt{liftershalte.info} collected waiting times hitchhiking ride this feature can already be found in the dataset during this phase.

According to the Internet Archive's Wayback Machine\footnote{\url{https://web.archive.org/web/20071201000000*/liftershalte.info}}\footnote{\url{https://web.archive.org/web/20080401000000*/maps.hitchwiki.org}} \texttt{liftershalte.info} was integrated into \texttt{maps.hitchwiki.org} between 2007 and 2008, until \texttt{liftershalte.info} ceased to exist around 2009. This is when \texttt{hitchwiki.org} took over the responsibility to maintain the leading online site to track and display hitchhiking via a map.

A significant turning point came in August 2010, when it was migrated to \texttt{hitchwiki.org/maps} with a visual makeover.\footnote{\url{https://web.archive.org/web/20100901000000*/hitchwiki.org/maps}} With this migration, recorded rides could be tied to a specific hitchhiker's profile, enabling users to document complete travel itineraries and facilitating a more comprehensive dataset. Beyond that, the collected information did not change much. However, via the project's code on GitHub\footnote{\url{https://github.com/PhilGruber/maps.hitchwiki.org}}, it can now be verified that waiting times were collected. Initially, contributions were modest, but in November 2010, the dataset saw rapid expansion thanks to a small group of power users. This period marked the high point of activity and usability, as the map built around the dataset matured into a practical tool for trip planning and research.

Although the original developer ceased active maintenance in 2012, the platform remained functional. However, this lack of oversight led to a gradual decline in stability and use. A sharp drop in user contributions occurred in November 2016, and by September 2020, the system had effectively broken, rendering it visible but non-functional for new data entries.

Due to the volunteering nature of the community around \texttt{hitchwiki.org} and the COVID-19 pandemic it took until October 2022, that a community member revived the platform by redeploying it as \texttt{hitchmap.com}\footnote{\url{https://hitchmap.com/}}, allowing new data to be entered once more. The platform has since been supported by a team of three developers, who expanded the feature set and continued maintenance. At this stage, free-form nicknames remained the primary method for tracking individual hitchhikers. By using their \texttt{hitchwiki.org} username as nickname some hitchhikers can still be tracked down over the different stages of the project.

Recent developments in 2025 include the reintroduction of user accounts, while also bridging \texttt{hitchmap.com} accounts,  to enhance the reliability of individual ride tracking and to enable the collection of richer metadata related to hitchhiking practices. The platform is once again experiencing growth, with ride submissions approaching levels that point towards what was last seen during its most active phase.

\begin{figure}[!h]
    \centering
     \includegraphics[width = 0.2\textwidth]{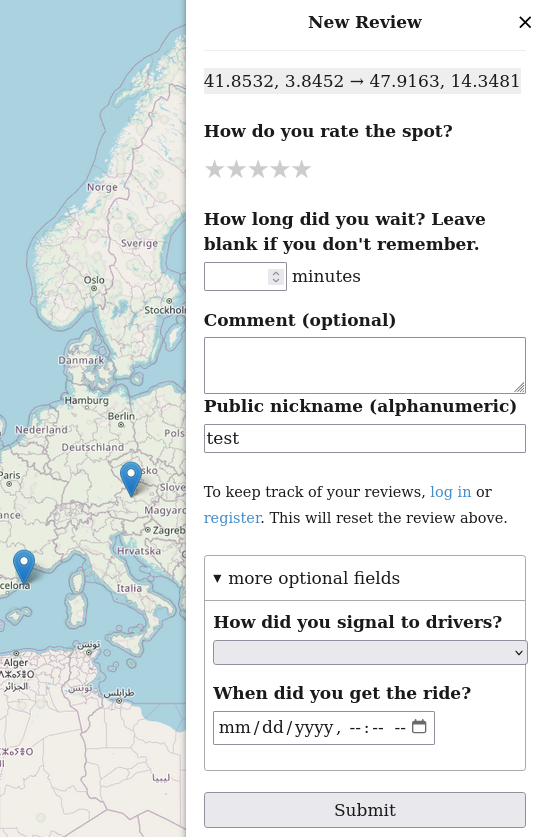}\\[1 ex]
    \caption{Form to input hitchhiking-ride-related information on \texttt{hitchmap.com}}
    \label{fig:hitchmap_form}
\end{figure}

\begin{table}
  \label{tab:hitchmap_form}
  \centering
  \begin{tabular}{ll}
    \toprule
    \cmidrule(r){1-2}
    \makecell{Feature prompt}     & \makecell{Input}     \\
    \midrule
    \makecell{Zoom the crosshair into your hitchhiking spot.\\Be as precisely as possible!}     & \makecell{[float, float]} \\
    \midrule
    \makecell{Where did your ride take you?\\Move the crosshair near the location, then press done.} & \makecell{[float, float]}\\
    \midrule
    \makecell{How do you rate the spot?} & \makecell{1, 2, 3, 4, 5 stars where 5 is best} \\
    \midrule
    \makecell{How long did you wait?\\Leave blank if you don't remember.}     & \makecell{int (minutes)}   \\
    \midrule
    \makecell{Comment (optional)}     & \makecell{string} \\
    \midrule
    \makecell{Public nickname (alphanumeric)}      & \makecell{string}       \\
    \midrule
    \makecell{How did you signal to drivers?}      & \makecell{[Sign, Thumb,\\Asked around,\\Asked around with a sign in hand]} \\
    \midrule
    \makecell{When did you get the ride?}      & \makecell{date-time}  \\
    \bottomrule
  \end{tabular}
  \caption{Prompts to ask for feature  and their possible input values on \texttt{hitchmap.com}}
\end{table}

At the time of writing \texttt{hitchmap.com} collects features of a hitchhiking ride through the form shown in \ref{fig:hitchmap_form} and fields presented in \ref{tab:hitchmap_form} could be collected.

The dataset reflects a dynamic and evolving platform history, shaped by user engagement, technological upkeep, and the commitment of a small but dedicated community. This background ensures that the dataset not only offers raw data but also mirrors the community-driven ethos of the hitchhiking movement.

\section{Dataset Composition}

The dataset captures the core spatiotemporal and experiential aspects that define the practice of hitchhiking. Each ride entry includes structured and semi-structured fields designed to reflect the complexities of hitchhiking as both a mode of travel and a cultural phenomenon.

At its core, every entry documents a starting location, identified by geographic coordinates (latitude and longitude), as interpreted via its location on Earth \footnote{\url{https://hitchwiki.org/en/Earth}}. This spatial anchoring has been part of the dataset since its early inception, providing the essential starting point of a hitchhiking ride.

A destination field was added in November 2022 following the transition to \texttt{hitchmap.com}. Rather than pinpointing a precise end location, this field captures a general direction of travel, reflecting the often fluid and open-ended nature of hitchhiking. This addition marked a key evolution in the dataset, allowing more comprehensive insights into route patterns and movement trends.

The temporal component of each ride is recorded through a submission date and time, which has been available since August 2010. Before that, only the date was stored. It is important to note that this timestamp does not necessarily reflect the actual date of the hitchhiking event, but rather the moment it was documented by the user.

Since  November 2024 an effort is made to also track the actual time at which the ride took place.

One of the dataset’s central purposes is to shed light on the waiting time — the duration a hitchhiker waited before being picked up. This metric is not only valuable for casual users but also for those interested in optimizing and understanding hitchhiking efficiency under varying conditions. As described in \ref{sec:methodology} this features must have been collected from the very start of this data record.

To enable ride tracking and trip reconstruction, each ride is associated with a nickname or user account if desired by the hitchhiker since August 2010. This user identifier allows rides to be grouped by individual hitchhikers and facilitates both community engagement and personal documentation. For many users, this transforms the platform into a place to record and reflect on their journeys — an experience shared with other platforms like \texttt{hitchlog.com}.\footnote{\url{https://www.hitchlog.com}}

Since March 2024, the dataset also includes a field documenting the method used to solicit a ride. Users can now specify whether they used a thumb (the classic method), a sign with a destination or direction, or approached drivers directly to ask for a lift. This addition allows for a deeper analysis of hitchhiking strategies and success rates.

Several subjective dimensions are also captured. Users can submit free-form comments, available since the beginning, to provide qualitative insights into their experience. While most comments are in English, they are not limited to it. Alongside these narratives, each ride can be assigned a rating, a feature available from the outset. This rating reflects the perceived quality of the spot and ride and links to a broader history within the \texttt{hitchwiki.org} community of quantifying hitchhiking conditions through the concept of "hitchability", which has been used to evaluate countries since 2010.\footnote{\url{https://hitchwiki.org/en/Hitchwiki:Hitchability}}

Together, these fields form a multifaceted dataset that not only records the logistics of hitchhiking trips but also captures the nuances of the hitchhiking experience. The dataset is released on Huggingface\footnote{\url{https://huggingface.co/datasets/Hitchwiki/hitchhiking-rides-dataset}} following the \textit{Hitchhiking Data Standard}\footnote{\url{https://github.com/Hitchwiki/hitchhiking_data_standard}} that was introduced to establish a common format which can help to collect better data on hitchhiking from multiple sources. Thus this work only caputures the current state of the published dataset. It is expected and desired that the dataset will grow by more entries from existing sources as well as records that are provided from new sources including privately collected data on hitchhiking and new online tools.

\section{Descriptive Analysis }\label{sec:analysis}

I want to provide an overview of the dataset presented in this work which will also grant insights into the world of hitchhiking. The code for generating those insights can be found on GitHub.\footnote{\url{https://github.com/Hitchwiki/hitchhiking-data/tree/8dac489b645d0e4e7e7d3ea82bebdf2834d95ee0/analysis/publications/2025\%20-\%20Wenke}}

\begin{figure}[!h]
    \centering
     \includegraphics[width=0.8\linewidth]{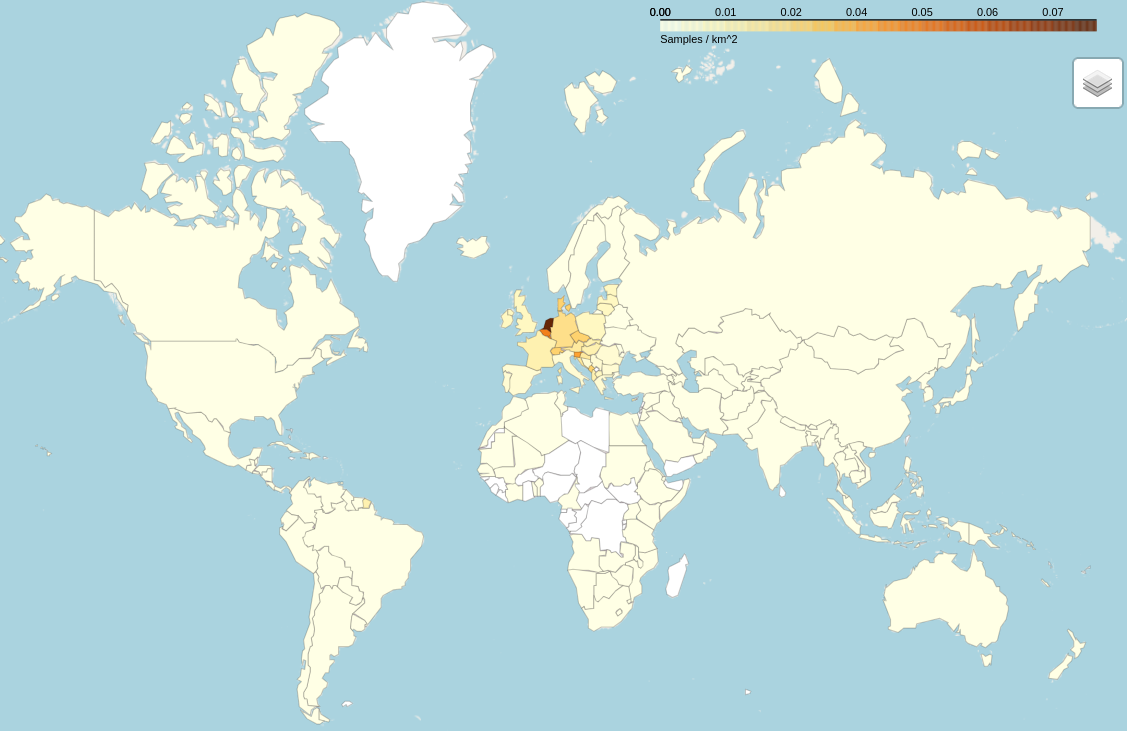}\\[1 ex]
    \caption{Rides per country relative to the country size}
    \label{fig:sample_density_by_country}
\end{figure}

First of all I acknowledge that the entire online hitchhiking community that centers around \texttt{hitchwiki.org} is very Europe-centric both in the content of articles as well as the authors of those articles. This can also be observed in \ref{fig:sample_density_by_country} where it is shown that rides that are recorded online concentrate on central Europe. It is to be investigated whether hitchhiking is as a phenomenon which is mostly practiced in Europe or by Europeans and in which forms it occurs in other parts of the world. 

\begin{figure}[!h]
    \centering
    \includegraphics[width=0.8\linewidth]{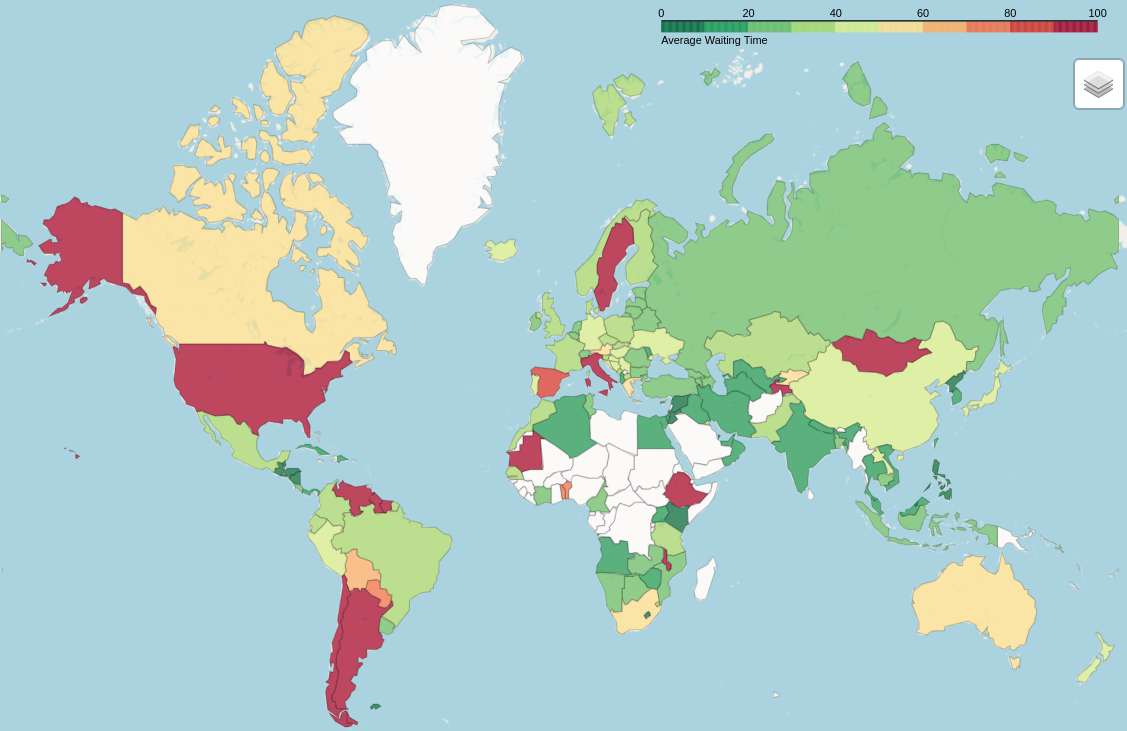}
    \caption{Average waiting time by country based on the submitted rides.}
    \label{fig:waiting_time_by_country}
\end{figure}

One of the most distinctive features of hitchhiking compared to other modes of transport is the inherent uncertainty in waiting time. This waiting time, defined as the duration from a hitchhiker beginning to look for a ride until they are picked up, serves as a core proxy for evaluating the quality and feasibility of hitchhiking in a given region. Figure~\ref{fig:waiting_time_by_country} presents the average waiting time by country. It becomes evident that reliability is closely tied to the density of recorded rides in the respective country or region.

Focusing on Europe, the dataset allows for a more fine-grained analysis of expected waiting times. This was explored in more detail in an earlier analysis from \cite{sulyok_average_2019}, which attracted considerable public interest.\footnote{\url{https://www.reddit.com/r/MapPorn/comments/bwf35p/hitchhiking_map_with_the_average_waiting_times_of/}} A global extension of this analysis was later published by \cite{wenke_heatchmap_2024}.

\begin{figure}[!h]
    \centering
    \includegraphics[width=0.8\linewidth]{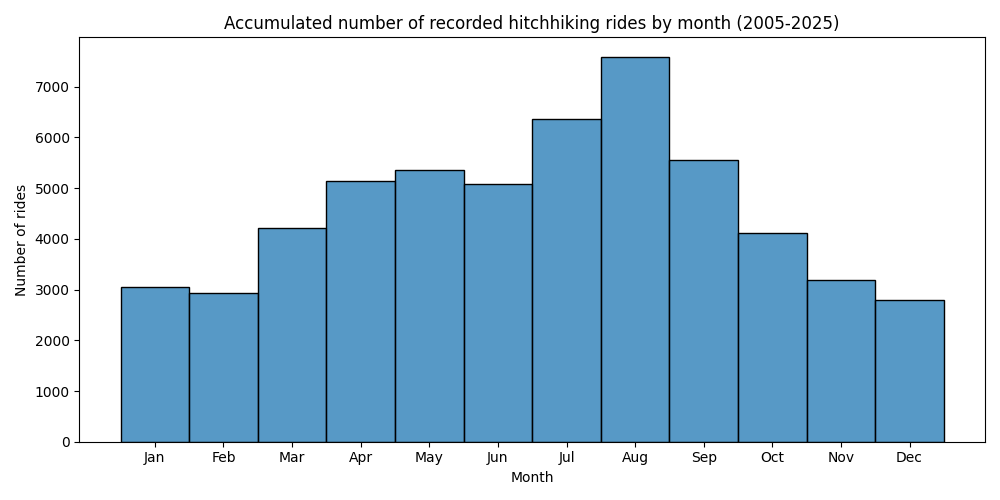}
    \caption{Number of recorded rides per month, showing seasonal trends.}
    \label{fig:number_of_rides_aggregated_by_month}
\end{figure}

Temporal dynamics in the dataset reveal a strong seasonality in hitchhiking activity. As previously suggested by Figure~\ref{fig:rides_per_month_2005_2025}, and further supported by Figure~\ref{fig:number_of_rides_aggregated_by_month}, ride submissions peak significantly during European summer months—roughly doubling in number compared to winter months. This seasonality likely reflects the weather dependence of hitchhiking. It is worth noting that the timestamps correspond to the moment a ride was recorded, which typically occurs shortly after the ride itself, possibly causing a slight lag in the peak months.

\begin{figure}[!h]
    \centering
    \includegraphics[width=0.8\linewidth]{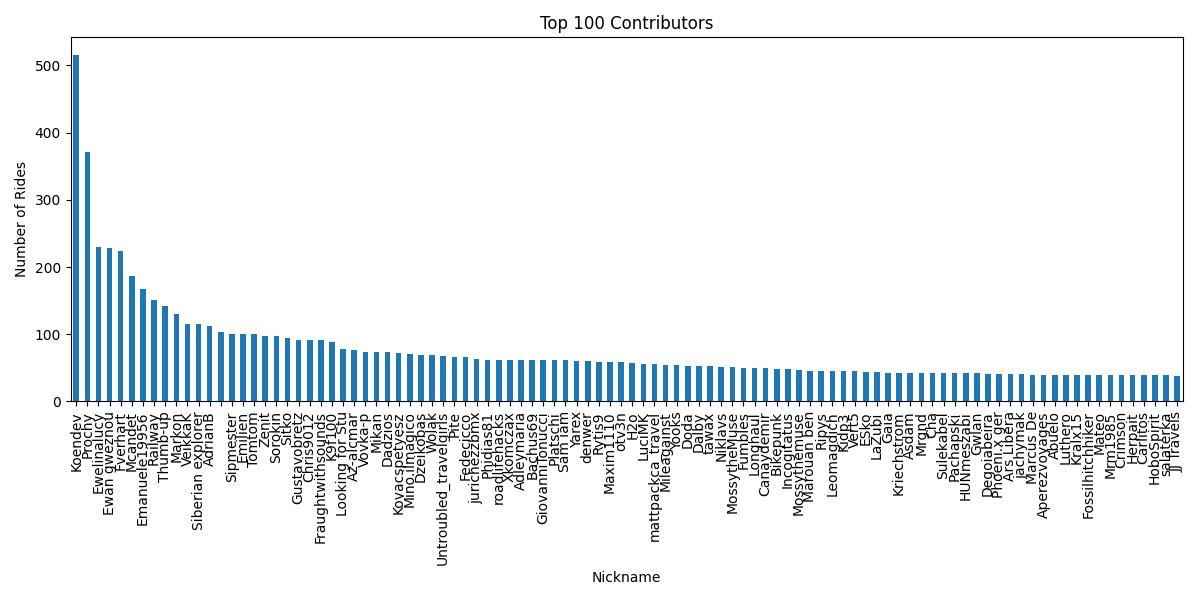}
    \caption{Top 100 contributors by number of submitted rides.}
    \label{fig:top_100_contributors}
\end{figure}

\begin{figure}[!h]
    \centering
    \includegraphics[width=0.8\linewidth]{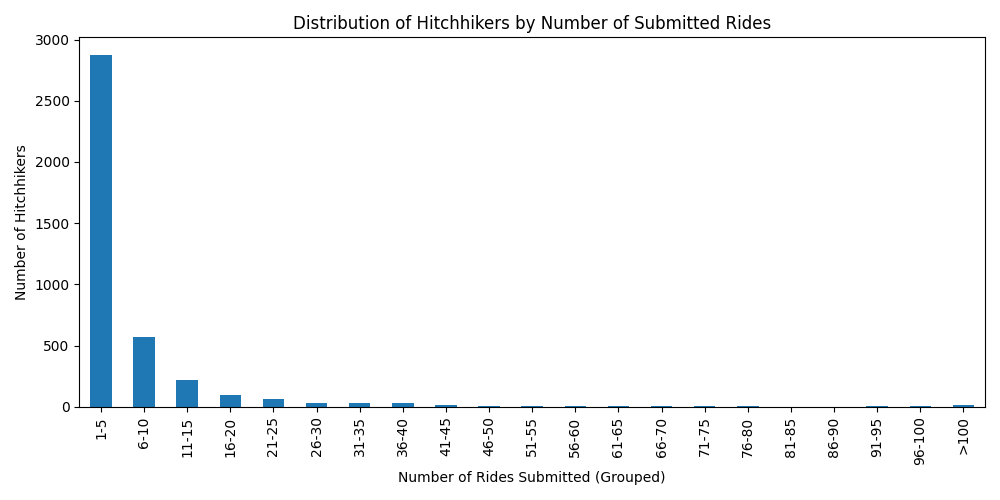}
    \caption{Distribution of hitchhikers by the number of submitted rides, exhibiting a power-law trend.}
    \label{fig:distribution_of_hitchhikers_by_number_of_submitted_rides}
\end{figure}

To assess the representativeness of the dataset, it is instructive to examine the distribution of contributions per user. An important question is whether the average hitchhiker's experience is well captured by the data, or whether the dataset is skewed by power users - those who record many rides and may differ in experience, skill or appearance from casual participants. Figures~\ref{fig:top_100_contributors} and~\ref{fig:distribution_of_hitchhikers_by_number_of_submitted_rides} show that the number of rides per user follows a power-law distribution. The vast majority of users have submitted fewer than five rides. This could suggest that most hitchhikers only hitchhike a few times in their life, perhaps during a special phase in their lives, or that many only hitchhike sporadically.

The presence of power users is often beneficial in the early stages of community growth, as their frequent contributions provide structure and value to the platform. However, it also introduces a potential bias in interpreting the dataset as representative of the broader hitchhiking population.

Finally, an analysis of ride comments offers additional insights. Approximately 70\% of the rides in the dataset include a user comment. Of these, 90\% are written in English, followed by 2\% in French, and 1\% in each Dutch and German. The remaining comments are written in various smaller European languages. This linguistic distribution further underlines both the dataset's European bias and the inherently international nature of hitchhiking.

\section{Limitations}\label{sec:limitations}

As the above section \ref{sec:analysis} indicates, this dataset does not constitute a representative sample of all hitchhiking activity globally. Estimating the overall proportion of global hitchhiking rides captured by the dataset remains an open and non-trivial challenge.

The dataset primarily reflects the activity of dedicated, long-term hitchhikers, many of whom are affiliated with the online community \texttt{hitchwiki.org}. As such, the contributor base is relatively homogeneous, limiting the diversity of perspectives and experiences captured. There is also a pronounced geographic bias towards Europe, both in terms of the origin of hitchhikers and the location of the rides.

Another key limitation is the inability to externally verify the authenticity of the submitted rides. The system relies on user honesty, and while various anti-spam measures have been introduced by the different collection-platforms over time, the risk of erroneous or fabricated entries cannot be entirely excluded. It also has to be noted that ride records that do not show a timestamp cannot be attributed to any source where they might have been collected, although it can be tracked down from their comments that they most probably have been collected from 2006 to 2010.

Beyond that, there is also information which was lost over the years. For example, this is true for \texttt{liftershalte.info} which already stored references to individual hitchhikers\footnote{\url{https://web.archive.org/web/20070615000000*/liftershalte.info}} that do not exist in this dataset anymore.

From a sociological perspective, the dataset is further limited by the lack of primary demographic and contextual data. No standardized collection of gender, age, driver characteristics, or stated intentions (of either hitchhikers or drivers) has been included. This restricts the dataset's utility for more comprehensive social science research on hitchhiking.

Efforts to address this gap have been initiated recently. In 2025, the \textit{Hitchhiking Data Standard} was proposed providing a foundation for decentralized and extensible data collection. This initiative aims to enable richer and more structured hitchhiking datasets in future iterations.

However, some limited demographic inference is possible. By cross-referencing user nicknames with accounts on \texttt{hitchwiki.org}, approximate gender data can be obtained for a subset of users. Based on this method, the contributing population is estimated to be approximately 82\% male and 18\% female. While this approach is imprecise and nonexhaustive, it reinforces the observation that hitchhiking, as represented in this dataset, remains a male-dominated activity.

\section{Outlook}

Looking ahead, there are several promising directions to expand and deepen this dataset. One important avenue involves integrating data from a broader range of sources, including privately maintained collections such as \url{https://prino.neocities.org}, and from platforms like \texttt{hitchlog.com}. Incorporating such external data could improve the dataset's geographic and demographic coverage, thereby allowing for a more representative view of global hitchhiking activity.

The relatively high proportion of rides accompanied by user comments suggests significant untapped potential for collecting richer information. These comments often contain details about context, experiences, and interpersonal dynamics, which are currently not captured in a structured format. This observation was a key motivation behind the development of the \textit{Hitchhiking Data Standard}, which aims to establish a flexible and extensible schema for future decentralized data collection efforts.

At present, the dataset provides only a limited, primarily technical perspective on hitchhiking, focusing on when, where, and how often rides occur. To develop a more holistic understanding of hitchhiking as a cultural practice, future work should also emphasize the interpersonal and social dimensions of the activity. This includes exploring the motivations, experiences, and relationships formed between hitchhikers and drivers, as well as the broader societal implications of hitchhiking as a mode of travel.

Finally, further research should investigate the global scope of hitchhiking. While current data is heavily biased toward Europe — and to a lesser extent, North America — it remains an open question to what extent hitchhiking is a truly global phenomenon, or whether it is largely constrained to specific cultural and geographic contexts. Addressing this question will require expanded outreach and collaboration with hitchhiking communities across the Global South and underrepresented regions.

\section{Safety and Ethical Discussion}\label{sec:ethical}

Hitchhikers often constitute a vulnerable population due to several inherent aspects of the practice: its legal ambiguity in some regions \footnote{\url{https://hitchwiki.org/en/Category:Illegal_to_hitchhike}}, the frequent reality of traveling alone, crossing national borders, entering remote or high-risk areas, and the uncertainty surrounding accommodation and shelter. These factors make safety a primary concern, not only during the act of hitchhiking but also when publicly sharing data related to such travels.

Individuals who choose to publish their trips online must carefully consider their personal safety. It is advisable to delay data publication slightly after the actual trip has occurred and to avoid personally identifiable details, such as using real names, exact home locations, or highly regular travel patterns. Thus most users adopt pseudonymous nicknames. Ultimately, however, the decision to publish and the level of detail shared remains a personal one.

Despite these risks, sharing ride data and travel logs can also offer significant value, both individually and collectively. Many hitchhikers document their journeys not only for personal reflection but to share experiences with a wider audience, often in blog or storytelling formats (e.g., \url{https://www.aenywhere.de}). These narratives help demystify hitchhiking, reduce stigma, and offer practical advice and encouragement to others considering alternative modes of travel.

Overall, this research and the associated data collection tools serve to support and promote hitchhiking as a legitimate and empowering form of travel. By making hitchhiking more visible, documented, and understood, such efforts contribute to keeping the practice alive in the modern world. Moreover, hitchhiking — when practiced thoughtfully — can help individuals overcome both physical borders and psychological boundaries, fostering connection, openness, and mutual trust across cultures and geographies.

\section{Conclusion}\label{sec:conclusion}

This paper has introduced and explored a unique, community-generated dataset capturing the practice of hitchhiking at an unprecedented scale. Comprising over 63,000 entries spanning nearly two decades, the dataset offers a rare lens into the geography, strategies, and temporal rhythms of hitchhiking across the globe, but particularly in Europe. Through a combination of historical tracing, descriptive statistics, and exploratory analysis, I have highlighted not only patterns in user behavior and waiting times but also the nature of participation in the online hitchhiking ecosystem.
My findings underscore the deeply situated and socially mediated nature of hitchhiking. The data reveals a mode of travel shaped by seasonal flows, regional norms, and the tacit knowledge of a relatively small yet highly engaged community. At the same time, the dataset reflects the limits of self-reported, crowd-sourced mobility data: geographic and demographic biases, uneven temporal coverage, and the unverifiable nature of individual reports all challenge standard notions of data objectivity and completeness.
Yet these limitations are also part of the dataset's value. As a living archive of situated experience, the hitchhiking rides data not only documents movement but also reflects an ethos of trust, spontaneity, and subversion of conventional transport systems. In this sense, the dataset is not simply a tool for analyzing hitchhiking; it is a cultural artifact of the practice itself.
Looking forward, this dataset opens multiple pathways for further research. These include comparative studies with other forms of informal or peer-to-peer mobility, deeper ethnographic or qualitative analysis of user contributions, and the integration of this dataset with other spatial or social data sources. As transportation systems become increasingly digitized and platform-mediated, studying grassroots, non-commercial mobility practices like hitchhiking becomes all the more important—for what they reveal not only about travel, but about genuine human connections.

\section{Acknowledgments}

The author would like to thank Philipp Gruber for operating and maintaining \texttt{liftershalte.info}, which played an essential role in hosting and sustaining early community infrastructure for hitchhiking data collection.

Special thanks also go to Kasper Souren, Mikael Korpela, and other contributors for founding and maintaining \texttt{hitchwiki.org}, with particular appreciation for the development and upkeep of \texttt{hitchwiki.org/maps}, which has been a foundational resource for the global hitchhiking community.

Finally, acknowledgment is due to Bob de Ruiter, Leon Weber, and the author for their collaborative efforts in continuing and expanding the work on \texttt{hitchmap.com}, a platform dedicated to visualizing and collecting hitchhiking ride data.

I want to especially thank Bob de Ruiter and Kasper Souren for reviewing a late stage of this work.

\section{References}

\bibliographystyle{plainnat}
\bibliography{main}

\begin{thebibliography}{9}
\providecommand{\natexlab}[1]{#1}
\providecommand{\url}[1]{\texttt{#1}}
\expandafter\ifx\csname urlstyle\endcsname\relax
  \providecommand{\doi}[1]{doi: #1}\else
  \providecommand{\doi}{doi: \begingroup \urlstyle{rm}\Url}\fi

\bibitem[Crassweller et~al.(1972)Crassweller, Gordon, and Tedford]{crassweller_experimental_1972}
Peter Crassweller, Mary~Alice Gordon, and W.~H. Tedford.
\newblock An {Experimental} {Investigation} of {Hitchhiking}.
\newblock \emph{The Journal of Psychology}, 82\penalty0 (1):\penalty0 43--47, September 1972.
\newblock ISSN 0022-3980, 1940-1019.
\newblock \doi{10.1080/00223980.1972.9916967}.
\newblock URL \url{http://www.tandfonline.com/doi/abs/10.1080/00223980.1972.9916967}.

\bibitem[Garrett(2014)]{garrett_hitchhiking_2014}
Mark Garrett.
\newblock Hitchhiking.
\newblock In \emph{Encyclopedia of {Transportation}: {Social} {Science} and {Policy}}. SAGE Publications, Inc., 2455 Teller Road, Thousand Oaks California 91320 United States, 2014.
\newblock ISBN 978-1-4522-6779-1 978-1-4833-4652-6.
\newblock \doi{10.4135/9781483346526.n271}.
\newblock URL \url{https://sk.sagepub.com/reference/encyclopedia-of-transportation/n271.xml}.

\bibitem[Guéguen*(2012)]{gueguen_color_2012}
Nicolas Guéguen*.
\newblock Color and women hitchhikers' attractiveness: {Gentlemen} drivers prefer red.
\newblock \emph{Color Research \& Application}, 37\penalty0 (1):\penalty0 76--78, February 2012.
\newblock ISSN 0361-2317, 1520-6378.
\newblock \doi{10.1002/col.20651}.
\newblock URL \url{https://onlinelibrary.wiley.com/doi/10.1002/col.20651}.

\bibitem[Kotz(2017)]{kotz_base-rate_2017}
Fabian Kotz.
\newblock The base-rate of hitch-hiking success and its moderators: {A} meta-analysis.
\newblock \emph{Transportation Research Part F: Traffic Psychology and Behaviour}, 46:\penalty0 149--160, April 2017.
\newblock ISSN 13698478.
\newblock \doi{10.1016/j.trf.2017.01.003}.
\newblock URL \url{https://linkinghub.elsevier.com/retrieve/pii/S1369847816300468}.

\bibitem[Sulyok(2019)]{sulyok_average_2019}
Abel Sulyok.
\newblock Average waiting times in {Europe}, June 2019.
\newblock URL \url{https://abelblogja.wordpress.com/average-waiting-times-in-europe/}.

\bibitem[Vedernikov et~al.(2016)Vedernikov, Kulik, and Ramamohanarao]{vedernikov_hitchhikers_2016}
Oleksii Vedernikov, Lars Kulik, and Kotagiri Ramamohanarao.
\newblock The {Hitchhiker}’s guide to the pick-up locations.
\newblock \emph{Open Geospatial Data, Software and Standards}, 1\penalty0 (1):\penalty0 12, December 2016.
\newblock ISSN 2363-7501.
\newblock \doi{10.1186/s40965-016-0012-1}.
\newblock URL \url{https://doi.org/10.1186/s40965-016-0012-1}.

\bibitem[Vedernikov et~al.(2017)Vedernikov, Kulik, and Ramamohanarao]{vedernikov_hitchhikers_2017}
Oleksii Vedernikov, Lars Kulik, and Kotagiri Ramamohanarao.
\newblock The {Hitchhiker}'s {Guide} to the {Optimal} {Route} {Planning}.
\newblock In \emph{2017 18th {IEEE} {International} {Conference} on {Mobile} {Data} {Management} ({MDM})}, pages 234--239, Daejeon, South Korea, May 2017. IEEE.
\newblock ISBN 978-1-5386-3932-0.
\newblock \doi{10.1109/MDM.2017.39}.
\newblock URL \url{http://ieeexplore.ieee.org/document/7962457/}.

\bibitem[Wenke(2024{\natexlab{a}})]{wenke_heatchmap_2024}
Till Wenke.
\newblock Heatchmap: {A} {Gaussian} process approach to predict hitchhiking waiting times, April 2024{\natexlab{a}}.
\newblock URL \url{https://tillwenke.github.io/2024/04/21/hitchmap-gp.html}.

\bibitem[Wenke(2024{\natexlab{b}})]{wenke_where_2024}
Till Wenke.
\newblock Where can {I} hitchhike?: {Hitchhiking} waiting times worldwide and per continent, May 2024{\natexlab{b}}.
\newblock URL \url{https://tillwenke.github.io/2024/05/06/hitchhiking-worldwide.html}.

\end{thebibliography}

\end{document}